# Amorphous intergranular films act as ultra-efficient point defect sinks during collision cascades

**Joseph E. Ludy[a], Timothy J. Rupert[a,b,*]**

[a] Department of Mechanical and Aerospace Engineering, University of California, Irvine, California, 92697, USA

[b] Department of Chemical Engineering and Materials Science, University of California, Irvine, California, 92697, USA

[*] Corresponding Author. Tel.: 949-824-4937; e-mail: trupert@uci.edu

**Abstract**

Atomistic simulations are used to explore the effect of interfacial structure on residual radiation damage. Specifically, an ordered grain boundary is compared to a disordered amorphous intergranular film, to investigate how interface thickness and free volume impacts point defect recombination. Collision cascades are simulated and residual point defect populations are analyzed as a function of boundary type and primary knock on atom energy. While ordered grain boundaries easily absorb interstitials, these interfaces are inefficient vacancy sinks. Alternatively, amorphous intergranular films act as ultra-efficient, unbiased defect sinks, providing a path for the creation of radiation-tolerant materials.

*Keywords:* grain boundaries; interface structure; molecular dynamics (MD); nanocomposite; amorphous intergranular films

The discovery of radiation-tolerant structural materials will play a vital role in the further development of nuclear energy technologies [1], as next-generation fission reactors and planned fusion technologies will require operation at higher radiation dosage rates. Inside a nuclear reactor, high-energy neutrons transfer kinetic energy to primary knock on atoms (PKAs), which then induce secondary collisions. These collision cascades create point defects within the material, with residual vacancies and interstitials eventually causing swelling and embrittlement that degrades mechanical behavior and leads to failure [2]. Historically, He ion bombardment has been used to replicate this process with lower PKA energies on the order of hundreds to thousands of electron volts. Experimental investigations of this type have shown that large He ion doses can cause defect planar clustering in Ni [3] and that a low angle grain boundary created by a network of edge dislocations can trap He in Mo at room temperature [4]. Recent research has suggested that the interfaces between crystallites (grain boundaries) can act as defect sink sites to help mitigate this problem [5], pointing to the promise of interface-dominated materials. For example, radiation-induced point defects are greatly reduced in layered Cu-V nanolaminates [6] and nanocrystalline Ni [7]. Similarly, nanostructured ferritic alloys [8] are possible candidates for radiation-resistant structures due to their larger interfacial volume fractions.

While prior work has shown that grain boundaries can increase the radiation tolerance of materials and nanostructuring has been pursued, the details of interfacial structure are also important for sink efficiency. For example, Demkowicz et al. found that the misfit dislocation networks that form at heterointerfaces in Cu-Nb multilayers can act as templates for point defect recombination [9]. Similarly, Samaras et al. demonstrated that areas of free volume within a boundary can act as recombination sites for point defects [10]. If free volume is an important factor for sink efficiency, one can hypothesize that an amorphous intergranular film (AIF) would be a promising structural feature for radiation tolerance, since amorphous materials such as metallic glasses have been shown to have an excess amount of free volume [11]. Crystalline materials with stable amorphous interfaces can now be produced by deposition techniques such as co-sputtering, where amorphous layers are created by solid state amorphization [12], or segregation engineering [13], where segregating dopant atoms can induce amorphization by lowering the energy penalty of having a glassy interfacial phase [14]. In this paper, we systematically compare collision cascade damage for a single crystal, an ordered grain boundary, and an amorphous intergranular film created by doping. Our results show that the AIF acts as an unbiased sink, absorbing both vacancies and interstitials, which is much more efficient than the ordered grain boundary.

The process of PKA damage occurs on the nanometer length scale over a few picoseconds, making atomistic simulations an ideal tool for studying such an event. Atomistic simulations capture the initial point defect distribution after the cascade stops but miss long-range diffusion effects [15], yet have been used extensively to provide insight into damage mechanisms [16]. In this paper, we use molecular dynamics (MD) run with the open-source Large-scale Atomic/Molecular Massively Parallel Simulator (LAMMPS) [17] code to simulate PKA damage in Cu and Cu-Zr. Atomic interactions are described by a potential from Mendelev et al. [18] which recreates important properties of both the elemental systems, using Embedded Atom Method (EAM) formulations, and glassy Cu-Zr phases, using a Finnis-Sinclair formalism. The short range forces of this potential are represented by a molecular statics method [19] that calculates the force on each atom as $F = F^{EAM} + \alpha F^g$, where $F^{EAM}$ is the force calculated from the EAM data in the potential and $\alpha F^g$ is the force contribution from a tested hybrid algorithm. Together, these terms account for the overestimation of repulsion at small separation distances.

An integration timestep of 1 fs is used for all simulations. The three test cases investigated here, shown in Figure 1, are a Cu single crystal, an undoped Σ5 (310) symmetric tilt grain boundary, and a 2.1 nm thick Cu-Zr amorphous intergranular film. In this figure, visualized by OVITO [20], crystalline Cu atoms are green, defect Cu atoms are white, and Zr atoms are black. These atom types were identified with adaptive common neighbor analysis [21] and all of our models consisted of ~108,000 atoms. The single crystal structure was used as a reference and had a [310] orientation with respect to the vertical axis of the simulation cell. A Σ5 (310) boundary was chosen to act as a model high-angle grain boundary, and was created by tilting both the top and bottom halves of a single crystal by 18.44° each about the [100] axis. To create the AIF, the Σ5 boundary structure was doped with 50 atomic % Zr in a 0.6 nm slice of material containing the boundary, heated locally to 1600 K to induce melting, and then cooled down to 300 K over 350 ps. This resulted in a 2.1 nm thick AIF with 15 atomic % Zr within the film.

Each simulation cell had periodic boundary conditions along all three axes to simulate a representative volume element of material. The models were equilibrated at 300 K and zero pressure under an isothermal-isobaric NPT (constant number of atoms, pressure, and temperature) ensemble for 10 ps. A PKA was then introduced at a distance of 0.5 nm from the interface with a velocity corresponding to 700, 1000, 1400, 2000, or 2500 eV of kinetic energy and directed upwards through the boundary, represented in Figure 1 as a red atom. The maximum PKA energy of 2500 eV was chosen because it is estimated to be the upper bound of what nanocomposites experience during He ion bombardment experiments [22]. The snapshots on the right in Figure 1 show the collision cascade of non-crystalline atoms at a PKA energy of 2000 eV when the ratio of non-crystalline to crystalline atoms is highest. The damage is relatively dispersed throughout both the single crystal and ordered boundary, but is localized closer to the AIF. Each structure was simulated five times using different seed values for the initial temperature, giving a total of 75 simulations. Changing the seed value for temperature slightly alters the development of the cascade event, allowing for a statistical analysis of residual damage.

For the next 52 ps, a microcanonical NVE (constant number of atoms, volume, and energy) ensemble was applied to the atoms in the cell interior where the collision cascade evolved. The boundary atoms were thermostatted at 300 K to simulate an infinitely large reservoir of room temperature material surrounding the cascade event, which forced the local temperature spike provided by the PKA to dissipate over time. Figure 2(a) shows the system-average temperature of the simulation cell, demonstrating the thermal spike induced by three select PKA energies in the model with an AIF. Figure 2(b) illustrates the evolution of atomic kinetic energy for a 2500 eV PKA at an AIF. The collision cascade begins at 0 picoseconds and the elevated temperature cools down over ~3 ps. At the beginning of this event, the energy is concentrated around the cascade core, which accounts for the temperature peak. While most atoms have kinetic energies on the order of 0.01 eV, the few higher energy atoms visible at 0 ps and 0.14 ps contribute greatly to the development of a kinetically 'hot' core and the large increase in the average temperature of the simulation cell. Most point defect recombination (~98%) occurred during this early period, but a small percentage of the defects recombine during the remainder of the equilibration simulation. The temperature peak occurs right when the PKA is introduced and the magnitude is proportional to the PKA energy. As time progresses, the number of hot atoms decreases until average temperature returns to 300 K.

Defect atoms, first identified by adaptive common neighbor analysis, were further analyzed using the Voronoi tool incorporated into LAMMPS [23]. Each atom was first assigned a Voronoi volume where any point in space is closer to that atom than any other. At the end of the simulation, the number of atoms inside each initial Voronoi volume is then calculated. If a pair of atoms is occupying a single Voronoi volume, it is an interstitial. Fig. 3 presents atomic snapshots of each model after the equilibration period has finished, with interstitials colored in red while defect atoms that retained one atom in their Voronoi volume are colored blue. In this figure, interstitials appear as a red atom surrounded by blue atoms, while vacancies appear as blue cells with the center atom missing. This tool does not explicitly find point defect types, but allows interstitials and vacancies to be manually counted by inspection according to the description above. For example, the Σ5 model shown in Fig. 3(b) has ten vacancies and three interstitials left in the simulation cell. As seen in Fig. 3(a), there are large numbers of residual interstitials and vacancies in the single crystal since direct recombination to eliminate Frenkel pairs is the only mechanism for defect annihilation. In the Σ5 model shown in Fig. 3(b), there are fewer interstitials remaining in the bulk because they have been absorbed by the boundary. However, similar to the observations of Bai and coworkers [16], many vacancies remain after equilibration. The AIF in Fig. 3(c) absorbs both interstitials and vacancies more efficiently, leaving the model sparsely damaged. Moreover, the remaining defects are localized near the boundary, as opposed to the Σ5 case where defects are found further away from the interface.

Figure 4 plots the average number of residual point defects as a function of PKA energy for the three simulation groups. As a general trend, the number of defects increases with increasing PKA energy for all models. The increase in residual damage with PKA energy is simply explained by the fact that larger numbers of point defects are created at higher energies. The ordered grain boundary outperforms the single crystal model in absorption of interstitials, but is not a consistently better vacancy sink. Alternatively, the AIF acts as an unbiased defect sink over all energies.

One reason for the improved sink efficiency of the AIF is its increased effective thickness. The cascade itself is far more contained within the 2.1 nm thick boundary when compared to the thin ordered grain boundary. Since the highest concentration of defects is produced near the center of the cascade [16], the defects are produced very close to where they can be absorbed. Caturla et al. used kinetic Monte Carlo simulations to show that large vacancy clusters form in the cascade core during an event [24]. This phenomenon can be taken advantage of with a thicker boundary, in terms of the kinetics of defect diffusion. Since the average mean free path for defect absorption will be shorter, more defects have the opportunity to be accommodated by a thicker boundary. While our current model does not account for recombination over longer time frames, a thicker boundary facilitates immediate defect absorption since damage is simply created closer to the sink.

The increased level of free volume in the AIF is likely another factor that leads to heightened point defect absorption. By again using an atomic Voronoi cell calculation, the volume associated with an average atom can be calculated for the single crystal, the ordered boundary, and the AIF. The average atomic volume associated with the single crystal can then be subtracted from the atomic volume within the ordered boundary and AIF to give a measure of free volume. While the ordered boundary only has a small amount of free volume when compared to the crystal (0.29 $\text{Å}^3$/atom), the AIF has ~4 times more free volume (1.16 $\text{Å}^3$/atom). This corresponds to a 2 percent and 10 percent increase in average atomic volume for the ordered boundary and AIF, respectively. It is well known that amorphous materials such as metallic glasses retain an excess amount of free volume due to their atomic packing [25]. Structural modeling efforts by Miracle [26] suggest that this excess may result from densely packed solute-centered atomic clusters that make up the metallic glass. In this model, there is no orientational order amongst these clusters and face-sharing of neighboring clusters is preferred to minimize solvent volume. Recently, Ma suggested that the presence of geometrically unfavored polyhedra frozen in during rapid quenching may contribute additional free volume to the overall structure [27]. Faster quenching rates create more readily configurable polyhedra, which increases the overall free volume. Tschopp et al. used MD to study how grain boundary character influences point defect formation in ordered grain boundaries, finding that self-interstitial atoms have a larger energetic driving force for binding to general, high-angle grain boundaries than vacancies do [28]. Since this explains why ordered interfaces struggle to absorb vacancies, the absence of such behavior in AIFs suggests that the added free volume in an AIF creates an additional driving force for vacancy binding. The addition of Zr dopants to the AIF may also contribute to this phenomenon. In Kirchheim's theoretical study of formation energies, segregated solutes within a boundary or near a dislocation were found to lower local vacancy formation energies [29].

Recent advances in processing science have made crystalline materials with amorphous interfaces accessible in multiple material systems. Magnetron sputtering is a viable processing method for the creation of such materials. Wang et al. used this method to fabricate Cu-Zr nanolaminates with alternating layers of crystalline Cu and amorphous Cu-Zr, created by solid state amorphization [12]. In this case, the thickness of the Cu-Zr amorphous layers could be tuned directly by controlling co-deposition times. However, this process is limited by low deposition rates and results in an anisotropic material, with the amorphous interfaces only being placed in the film growth direction. The formation of disordered interfacial complexions via segregation engineering [30] is a promising technique for introducing AIFs without such limitations. Shi and Luo developed the thermodynamic theory behind such disordered films, showing that doping can reduce the free energy penalty for the formation of AIFs and creating predictive grain boundary diagrams [14]. Tang et al. also developed a thermodynamic model incorporating interface energy and temperature [31], showing that disordered complexions are more stable at higher temperatures and that high-angle boundaries provide a more stable site for complexions than low-angle grain boundaries. An implication of these models is that AIFs may be introduced into a random polycrystalline material if segregating dopants are added and high temperature grain boundary structures are quenched into the microstructure, with thickness perhaps controlled by tuning dopant concentration and temperature.

Our molecular dynamics simulations suggest that amorphous intergranular films act as efficient point defect sinks when compared to general, high-angle grain boundaries. Increasing the effective interfacial thickness allows for shorter point defect migration distances, while the excess free volume present in an AIF dramatically increases the absorption of vacancies. The extremely rapid healing of cascade damage at an AIF should be beneficial for nuclear reactor materials which must survive high dosage rates. Incorporating AIFs into nanostructured materials would place these damage-tolerant interfaces regularly throughout the material, offering a promising route for unprecedented resistance to radiation damage.

This research was supported by the U.S. Department of Energy, Office of Basic Energy Sciences, Materials Science and Engineering Division under Award No. DE-SC0014232. T.J.R. acknowledges support from a Hellman Fellowship Award.

**References**

[1] S.J. Zinkle, G. Was, Acta Materialia, 61 (2013) 735-758.

[2] K.L. Murty, I. Charit, Journal of Nuclear Materials, 383 (2008) 189-195.

[3] A. Van Veen, J. Evans, L. Caspers, J.T.M. De Hosson, Journal of Nuclear Materials, 122 (1984) 560-564.

[4] J. Evans, A. Van Veen, J.T.M. De Hosson, R. Bullough, J. Willis, Journal of Nuclear Materials, 125 (1984) 298-303.

[5] W. Han, M.J. Demkowicz, N.A. Mara, E. Fu, S. Sinha, A.D. Rollett, Y. Wang, J.S. Carpenter, I.J. Beyerlein, A. Misra, Advanced Materials, 25 (2013) 6975-6979.

[6] E. Fu, J. Carter, G. Swadener, A. Misra, L. Shao, H. Wang, X. Zhang, Journal of Nuclear Materials, 385 (2009) 629-632.

[7] N. Nita, R. Schaeublin, M. Victoria, R. Valiev, Philosophical Magazine, 85 (2005) 723-735.

[8] M. Demkowicz, P. Bellon, B. Wirth, MRS Bulletin, 35 (2010) 992-998.

[9] M. Demkowicz, R. Hoagland, J. Hirth, Physical Review Letters, 100 (2008) 136102.

[10] M. Samaras, P. Derlet, H. Van Swygenhoven, M. Victoria, Philosophical Magazine, 83 (2003) 3599-3607.

[11] A.R. Yavari, A. Le Moulec, A. Inoue, N. Nishiyama, N. Lupu, E. Matsubara, W.J. Botta, G. Vaughan, M. Di Michiel, Å. Kvick, Acta Materialia, 53 (2005) 1611-1619.

[12] Y. Wang, J. Li, A.V. Hamza, T.W. Barbee, Proceedings of the National Academy of Sciences, 104 (2007) 11155-11160.

[13] D. Raabe, M. Herbig, S. Sandlöbes, Y. Li, D. Tytko, M. Kuzmina, D. Ponge, P. Choi, Current Opinion in Solid State and Materials Science, 18 (2014) 253-261.

[14] X. Shi, J. Luo, Physical Review B, 84 (2011) 014105.

[15] W. Phythian, R. Stoller, A. Foreman, A. Calder, D. Bacon, Journal of Nuclear Materials, 223 (1995) 245-261.

[16] X.M. Bai, A.F. Voter, R.G. Hoagland, M. Nastasi, B.P. Uberuaga, Science, 327 (2010) 1631-1634.

[17] S. Plimpton, Journal of Computational Physics, 117 (1995) 1-19.

[18] M. Mendelev, M. Kramer, R. Ott, D. Sordelet, D. Yagodin, P. Popel, Philosophical Magazine, 89 (2009) 967-987.

[19] M. Mendelev, D. Sordelet, M. Kramer, Journal of Applied Physics, 102 (2007) 043501.

[20] A. Stukowski, Modelling and Simulation in Materials Science and Engineering, 18 (2010) 015012.

[21] A. Stukowski, Modelling and Simulation in Materials Science and Engineering, 20 (2012) 045021.

[22] M. Demkowicz, R. Hoagland, International Journal of Applied Mechanics, 1 (2009) 421-442.

[23] C.H. Rycroft, Chaos, 19 (2009) 041111.

[24] M. Caturla, N. Soneda, E. Alonso, B. Wirth, T.D. de la Rubia, J. Perlado, Journal of Nuclear Materials, 276 (2000) 13-21.

[25] A. Slipenyuk, J. Eckert, Scripta Materialia, 50 (2004) 39-44.

[26] D.B. Miracle, Nature Materials, 3 (2004) 697-702.

[27] E. Ma, Nature Materials, 14 (2015) 547-552.

[28] M.A. Tschopp, K. Solanki, F. Gao, X. Sun, M.A. Khaleel, M. Horstemeyer, Physical Review B, 85 (2012) 064108.

[29] R. Kirchheim, Acta Materialia, 55 (2007) 5129-5138.

[30] P.R. Cantwell, M. Tang, S.J. Dillon, J. Luo, G.S. Rohrer, M.P. Harmer, Acta Materialia, 62 (2014) 1-48.

[31] M. Tang, W.C. Carter, R.M. Cannon, Physical Review B, 73 (2006) 024102.

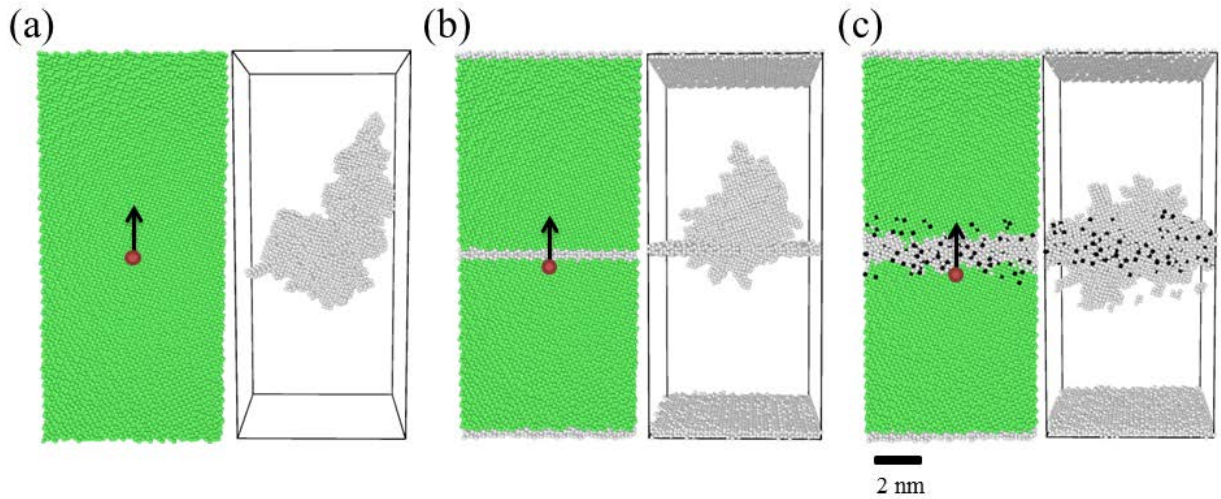

**Figure 1: Collision cascade snapshots before (left) and during (right) cascade events of the (a) single crystal, (b) Σ5 (310), and (c) AIF structures with a PKA energy of 2000 eV. The PKA is colored red and directed upwards 0.5 nm away from each boundary. The green atoms are of the face centered cubic structure type and the white atoms are defects. The black atoms are Zr dopants.**

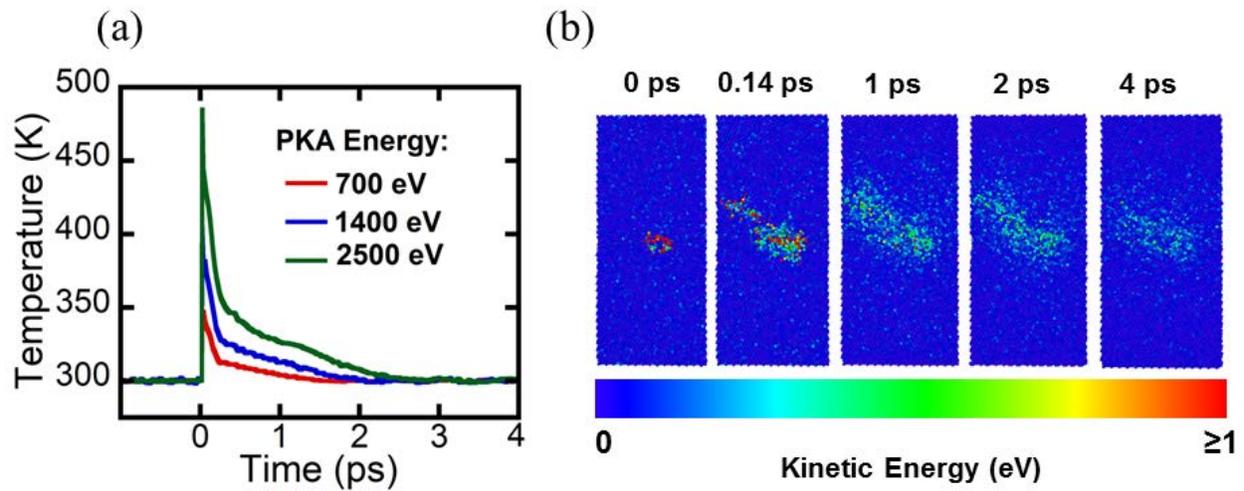

**Figure 2: (a) The average temperature of the simulation cell in the AIF model during the 700, 1400, and 2500 eV PKA collision cascades. (b) The atomic kinetic energy distribution during a 2500 eV cascade. The cascades begin at 0 ps and the system cools down to 300 K over ~3 ps.**

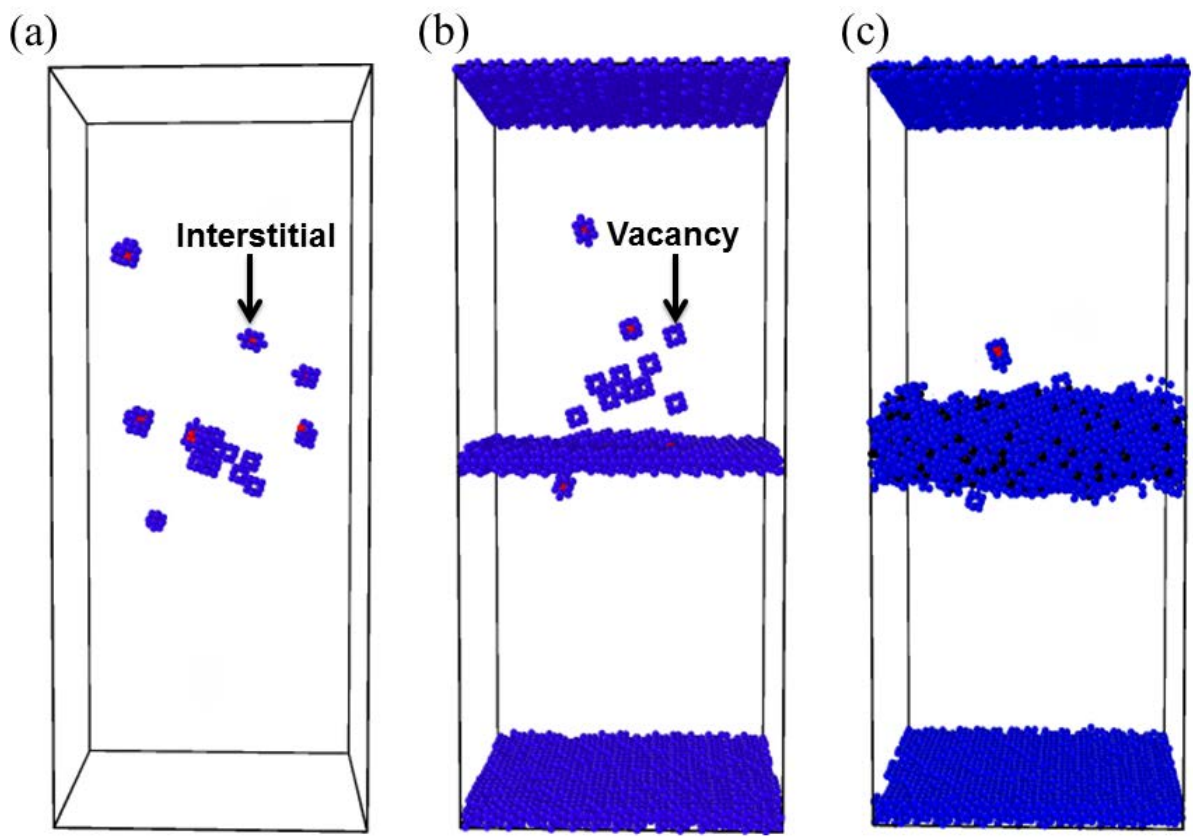

**Figure 3:** Post-equilibration snapshots of the (a) single crystal, (b) Σ5 (310) boundary, and (c) AIF models after a 2000 eV PKA. Voronoi volumes containing more than one atom are colored red (interstitials) and Voronoi volumes containing only one atom are colored blue. Zr atoms are colored black.

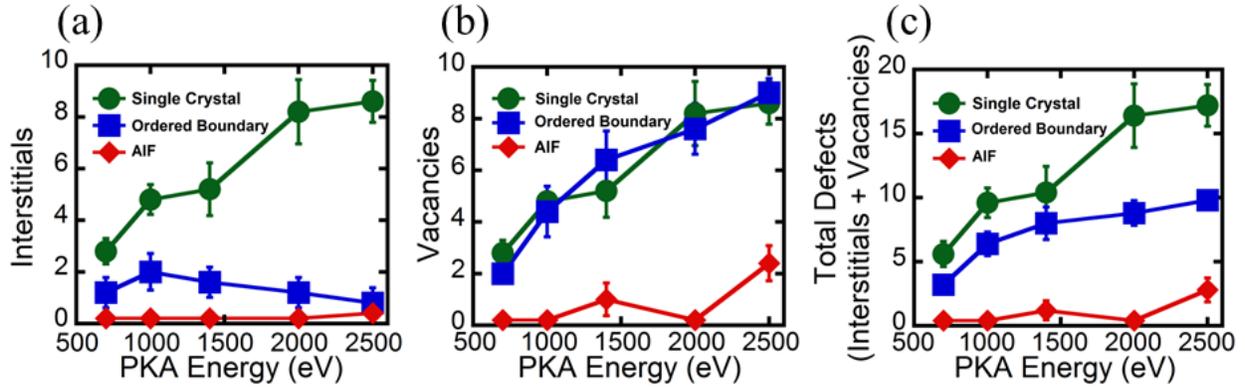

**Figure 4:** The number of residual (a) interstitials, (b) vacancies, and (c) total defects as a function of PKA energy. The Σ5 (310) boundary preferentially absorbs interstitials, but the AIF acts as a superior, unbiased sink.